# Sleep Paralysis: phenomenology, neurophysiology and treatment

Elizaveta Solomonova[1,2]

[1]Université de Montréal, Individualized program (Cognitive Neuroscience & Philosophy).
[2]Center for Advanced Research in Sleep Medicine, Dream and Nightmare Laboratory, Montreal, Canada

## Abstract

Sleep paralysis is an experience of being temporarily unable to move or talk during the transitional periods between sleep and wakefulness: at sleep onset or upon awakening. Feeling of paralysis may be accompanied by a variety of vivid and intense sensory experiences, including mentation in visual, auditory, and tactile modalities, as well as a distinct feeling of presence. This chapter discusses a variety of sleep paralysis experiences from the perspective of enactive cognition and cultural neurophenomenology. Current knowledge of neurophysiology and associated conditions is presented, and some techniques for coping with sleep paralysis are proposed. As an experience characterized by a hybrid state of dreaming and waking, sleep paralysis offers a unique window into phenomenology of spontaneous thought in sleep.

## Introduction

*"I had a few terrifying experiences a few years ago. I awoke in the middle of the night. I was sleeping on my back, and couldn't move, but I had the sensation I could see around my room. There was a terrifying figure looming over me. Almost pressing on me. The best way I could describe it was that it was made of shadows. A deep rumbling or buzzing sound was present. It felt like I was in the presence of evil... Which sounds so strange to say!"* (31 year old man, USA)

Sleep paralysis (SP) is a transient and generally benign phenomenon occurring at sleep onset or upon awakening. Classified as a rapid-eye-movement (REM) sleep-related parasomnia, SP represents a psychophysiological state characterized simultaneously by qualities of both sleep and wakefulness, wherein the experiencer can open her eyes (Hishikawa & Kaneko, 1965), can be aware of her physical environment but is unable to move and may start seeing, hearing, feeling or sensing something.

While documented instances of SP seem to be very consistent across cultures , SP's lived qualities, phenomenology, and interpretation as a meaningful experience varies depending on the cultural and religious background. Rooting SP in a particular belief system may either help the experiencer recognize that SP is common and transient, or amplify negative qualities of SP by giving more concrete shape to an already terrifying experience of a supernatural assault.

The emotional experience of SP is often one of fear, terror and panic. Threatening presences, the vulnerability of being in a paralyzed state, uncontrollable visions - all these elements contribute to intense, predominantly dysphoric, negative affect. Some qualities of spontaneous thought associated with felt presence during SP can be seen as paranoid (Cheyne & Girard, 2007), spatial, or interpersonal/social (Nielsen, 2007, Solomonova, 2008). The vast majority of SP experiences are associated with intense feelings of realism, and are most often characterized by fear and distress, which may carry-over into wakefulness and create a vicious cycle of negative emotional association with sleep, including aversion to going to bed, and even, in extreme cases, can result in symptoms reminiscent of post-traumatic stress disorder (McNally and Clancy, 2005). Yet, some SP experiences are described in positive terms, especially vestibulo-motor phenomena that include out-of-body experiences (OBE), or sensations of flying or floating. While the intuitive and immediate reaction to SP is typically negative, as will be discussed in the section on practical considerations, there are numerous reports of neutral/positive SP. Furthermore, there is a possibility of harnessing the power and potential of the dissociative/overlapping state in order to take active charge of one's experience and to use the opportunity presented by the simultaneity of waking and sleeping cognition - a potential for entering into a lucid dream state or for contemplative self-observation.

Neurophysiologically, SP is currently understood as a state dissociation or a state overlap between REM sleep and wakefulness (American Academy of Sleep Medicine, 2014). During SP one can open her eyes, look around the room, become aware of her environment and simultaneously experience REM sleep-related paralysis (muscle atonia) as well as intense and realistic imagery[1] of all sensory modalities - a nightmare spilling into the real world. Normally, during REM sleep, skeletal muscle atonia blocks most motor output, effectively preventing the sleeper from acting out her dreams (Peever, Luppi, & Montplaisir, 2014). SP can also occur in the context of narcolepsy (Sharpless & Barber, 2011; Terzaghi, et al, 2012), but the majority of those who experience SP report it in its isolated form (often referred to as Isolated Sleep Paralysis), without known medical or neurological association.

In current medical and neuroscientific literature, SP is discussed in terms of its presentation and negative factors: SP-associated mentation is generally seen as a non-desirable effect of REM sleep intrusion into waking. In this chapter I propose that situating SP experiences as a dream phenomenon within the framework of embodied mind and enactive cognitive science, including contemplative approaches to consciousness, is an alternative that accounts for the phenomenology of SP as a lived experience, allows for rich and detailed cultural framing of the experience, and offers avenues for a cross-cultural social neurophenomenology of SP.

First, I will discuss the phenomenology and neurophysiology of SP experiences. I will present SP in general, without distinguishing between its isolated and narcolepsy-related form, unless such a separation is warranted. I will start by presenting the current state of knowledge of SP prevalence, as well as the varieties of imagery accompanying SP and their cultural significance. I will then discuss SP in terms of a REM sleep parasomnia and outline its precipitating and enabling factors, as well as sleep and dream characteristics of those who experience SP. Finally, I will examine SP in light of various cultural and shared practices, including preventative measures, and practices aimed at interrupting and transforming SP experiences.

The experiential examples of SP used in the present chapter are all derived from an Internet-based study of SP (Solomonova et al, 2008). Our research group collected 193 responses from people with recurrent SP experiences, using a modified version of the Waterloo Unusual Sleep Experiences Questionnaire (Cheyne, Newby-Clark, & Rueffer, 1999). Our participants were recruited online, using word of mouth, and via advertising on SP-related forums, information and support.

### Definitions and prevalence

Idiopathic SP (SP not associated with narcolepsy, and without known cause) is a benign and transient parasomnia (Howell, 2012), occurring during transitions between wake and sleep: at sleep onset or upon awakening. The Diagnostic and Statistical Manual of Mental Disorders (DSM-5) classifies isolated SP accompanied by fearful mentation as an instance of a nightmare disorder (American Psychiatric Association, 2013). During an episode of SP, characteristics of REM sleep intrude upon seemingly awake consciousness: thus the person experiencing SP, while having an impression of being awake and aware of her environment, is unable to initiate voluntary movements (i.e., experiences REM sleep muscle atonia/paralysis), and may also experience intense and realistic sensations in any sensory modality - REM sleep-related mentation (American Academy of Sleep Medicine, 2014). SP should be distinguished from night terrors – early night awakenings with feelings of panic/terror, typically associated with somnambulism-spectrum arousal from slow wave (Stages 3 and 4) non-REM (NREM) sleep. Night terrors are characterized by sudden awakening in an agitated state, anxiety, body motility and general amnesia with regards to underlying cognitive experience (Szelenberger, Niemcewicz & Dabrowska, 2009).

Prevalence estimates of SP range widely, and may depend on geographic and cultural factors. A systematic review of SP prevalence has revealed that studies report SP lifetime prevalence from as low as 1.5% to possibly 100% in the general population (Sharpless & Barber, 2011). The authors indicated that about one in five individuals may have experienced SP at least once in their lifetime (of 36,533 persons in their review). Prevalence estimation of SP is difficult due to numerous factors such as ethnicity and cultural background, including variable familiarity with the phenomenon and in the wording of questions (Fukuda, 1993). For example, in a cross-cultural study, Fukuda and colleagues (2000) reported that while it is unclear whether SP is equally prevalent in Canada and Japan, the lack of familiarity and of a normative cultural framework for SP in Canada may contribute to the fact that many Canadian, but not Japanese, respondents qualified SP as a kind of a dream, and would not have, therefore, readily recognized SP in a prevalence study. An additional reason for under-diagnosis of SP in the West may be the fact that those who experience SP may be misdiagnosed as having psychiatric disturbances (Hufford, 2005). The developmental trajectory of SP is traditionally associated with an onset during adolescence, which may indicate a process associated with sleep architecture maturation (Wing, Lee & Chen, 1994). However, in one study of older adults, a bimodal onset pattern was reported, with a second pattern of onset of SP episodes after the age of 60 years old (Wing et al, 1999), suggesting a possibility that SP may have a variety of onset conditions.

---

[1] In this chapter I will use the terms 'imagery' and 'mentation' interchangeably to refer to visual, auditory, somatosensory and even social experiences during SP. The term 'imagery' here is therefore not restricted to the visual domain. I prefer 'imagery' and 'mentation' to 'hallucination' in order to emphasize the dream-like process of spontaneous imagination that takes place during SP, and to de-emphasize the association with delusional thought and pathologies, associated with the term 'hallucination'.

Neurologically and phenomenologically SP is situated on the REM-sleep based dream/nightmare continuum. In this chapter, I will approach SP experiences as a variant of intensified or disturbed dreaming, and will situate them within a framework of embodiment and enactivism.

### *The 4EA cognition and oneiric mentation*

Recent years have seen the development of a paradigm shift from a strictly neurocentric view of the mind, a position that can be stated as "embrained" (Morris, 2010), to a diverse family of approaches that consider the mind embodied (Varela, Thompson & Rosch, 1991; Gallagher, 2005), enactive (Noe, 2004; Thompson, 2005; Stewart, Gapenne & Di Paolo, 2010), extended into and embedded in the physical and social world (Clark & Chalmers, 1998; Menary, 2010), and affective (Colombetti, 2013; Pessoa, 2013). While these approaches are in many respects quite different, they have sometimes been labeled as 4EA (embodied, embedded, extended, enactive, affective) cognition, with a common theme of offering a robust alternative to computational, connectionist, and neuro-reductionist views of the mind (Wheeler, 2005; Protevi, 2012). These theories attempt to situate cognition, brain activity, and psychophysiology within the larger contexts of lived subjective experience, by emphasizing the roles of developmental sensorimotor attunement to the world, as well as of the active and motivated processes of perception and sense-making, the importance of the social and cultural milieu, and the role of emotion and affect.

Sleep and dreaming phenomena have been only rarely addressed by 4EA theorists (with the exception of Thompson 2014, 2015a,b), and the prevailing view of the sleeping mind today situates sleep mentation as being firmly constrained within the brain (Rechtschaffen, 1978; Hobson, Pace-Schott & Stickgold, 2000; Revonsuo, 2006). As Revonsuo states: "The conscious experiences we have during dreaming are isolated from behavioral and perceptual interactions with the environment, which refutes any theory that states that organism-environment interaction or other external relationships are constitutive of the existence of consciousness" (Revonsuo et al, 2015: 3). Alternatively, situating dream mentation within a framework of 4EA approaches implies that the dreaming subject is not entirely isolated or disconnected from environmental and somatic stimuli, and that her experiential self retains affective, social, sensorimotor and sense-making qualities. Dreaming then is not passively lived as a purely mental simulation (Revonsuo et al, 2015), but can be seen as a process of active imagination (Thompson, 2014) rooted in the dreamer's physical, social and affective world (Solomonova & Sha, 2016). I propose that SP experiences, by virtue of their special kind of overlap between and simultaneous presence of both waking and dreaming cognition, are perfect candidates for neurophenomenological research on spontaneous thought in sleep, which would help illuminate particular qualities of dreaming cognition that may otherwise be inaccessible to reflective consciousness upon awakening from a dream.

***Phenomenology of sleep paralysis experiences***

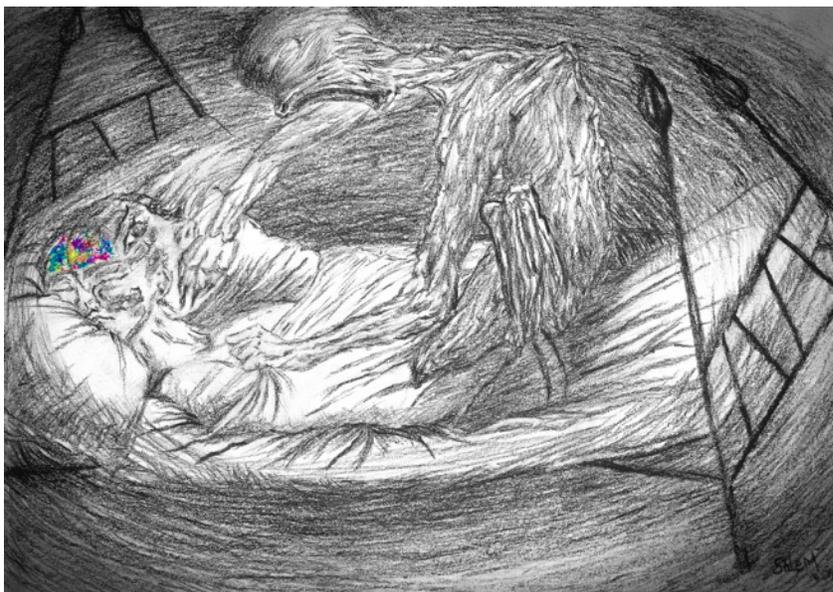

**Figure 1.** A representation of sleep paralysis experience[2]. The sleeper is awakened suddenly and sees a menacing shadowy creature on top of him. He experiences the sensation of being pushed into the bed, while the bed itself is swirling in a sort of a tornado. The two faces of the dreamer represent the "double" consciousness during sleep paralysis: he is simultaneously terrified of the supernatural attacker and also knows that if he does not resist the experience and allows himself to drift back into sleep he may have a lucid dream (this lucid consciousness is represented as a sleeping face with a colorful brain, denoting vibrant possibilities of lucidity. Artist: Benjamen Samaha, Montreal, Canada 2016. Reproduced with artist's permission.

In addition to transient experience of muscle paralysis, the most dramatic quality of SP is its sensory content, characterized by vivid, intrusive audio-visual and somatosensory imagery. The experience of SP can be extremely realistic, have a quasi-perceptual and wake-like quality, and may be accompanied by tactile and kinesthetic sensations. Reflective thought processes, self-awareness and metacognitive abilities seem to be relatively preserved during SP experiences, and people who have had multiple SP experiences may develop a "feel" for recognizing SP imagery.

SP-associated experiences are typically referred to as hallucinations (hypnagogic, when occurring at sleep onset, or hypnopompic, when happening upon awakening), since these occur during otherwise seemingly awake consciousness (Liddon, 1967; American Academy of Sleep Medicine, 2014). This entails that a person who experiences SP sees something that is not there, something that is distorted or false. Such a view presupposes that during SP one is effectively awake and is misinterpreting her experience. Another way of looking at SP is to situate it within the spectrum of dream mentation and dreaming imagination. While dreaming too has been seen as a delusional/hallucinatory activity, an alternative view, in line with embodied mind theories and enactive approach, has also been proposed: "When you hallucinate, you seem to perceive what is not there. (...) When you imagine, you evoke something absent and make it mentally present to your attention" (Thompson, 2014: 179). In this chapter I adopt this latter view and will refer to SP experiences as variant of spontaneous thought/mentation or mental imagery, rather than hallucinations or delusions.

---

[2] Excerpt from the dreamer's account: "The transition between wake and sleep is a crucial moment to enter into the world of dreams. During this transition on countless occasions I would awaken suddenly not being able to move. During this experience it seems that the very essence of fear permeates my consciousness. Eeriness goes through my soul, freezes my blood and interrupts all substantial notion of my being. No words can describe that visceral sensation, and in parallel, no words can come out of my mouth. Aware of the lack of muscle tonus, I try to escape this inevitable Machiavellian *black beast*, that materializes in my head before my eyes and on my chest, slowing down my breathing. Moreover, my senses are grabbed by an impression of fighting a hurricane that may drag me out of my body. This cyclone, that has a black hole in lieu of an eye, forces me to fight it, and this fight seems crucial to my survival (...) In addition, there are auditory experiences, an amalgam of petrifying words and vibrations that feel like sudden gusts of wind in the eardrums. All this happens when by body feels like a statue, without a possibility of screaming.
(...) On occasion, with determination and *lucidity*, I can have power over this swirl of stillness. (...) I take back some control of my imaginary hands, and then I hold them out to *Morpheus* for a dazzling and colorful dance in a deep and enlightened night". (Male SP sufferer, also diagnosed with narcolepsy. Montreal, Canada. Account translated from French).

## Kinds of sleep paralysis experience

A factor analysis by Cheyne and colleagues (1999) showed that SP mentation typically falls into three general categories. The first category is *Intruder,* and it is characterized by a felt presence, fear, as well as auditory and visual imagery. The person who experiences SP feels that someone is in the house or in their room. This experience is sometimes accompanied by seeing or hearing someone or something sentient move around the house. The second category of SP experiences is known as *Incubus*, in which the felt presence is interpreted as a supernatural assault and is often accompanied by a sensation of shallow breathing, a feeling of being smothered, pressure on the chest, or pain. In this case, the sleeper often sees and feels the maleficent being on top of her. The third category, *Unusual Bodily Experiences*, appears to be a separate, less well-known, and a qualitatively different kind of SP experience: these are often described as positive events, such as sensations of floating, out-of-body experiences (OBEs), and feelings of bliss. Both *Intruder* and *Incubus* categories typically include the experience of felt presence – a distinct sensation that someone sentient is in the immediate vicinity of the sleeper (Cheyne, 2005).

Most literature on SP focuses almost exclusively on the first two kinds, *Intruder* and *Incubus,* possibly due to their particularly intensified felt presence imagery, which contributes to distressing SP experiences (Solomonova et al, 2008; Cheyne, 2013). However, neutral and positive instances of SP have also been described, and the third category, *Unusual Body Experiences*, or vestibulo-motor experiences, is often characterized by pleasant sensations and a spirit of exploration, accompanying sensations of flying, out-of-body experiences, or autoscopy (observation of one's own body from an unusual/novel point of view) (Brugger, Regard & Landis, 1997).

## Felt Presence

*"Just before going to sleep or if awoken suddenly I feel as though a presence, usually a dark shadow figure is standing over the bed staring down at me, or pacing back and forth."* (22 year old, gender not reported, USA).

Among all SP experiences, felt presence, the distinct sensation that another sentient being, human or not, is present in the extracorporeal space of the experiencer, is arguably the most salient, terrifying, and rich.. Felt presence is consistently reported as the most common SP-associated experience – about 80% of episodes (Cheyne et al, 1999), which produces most fear and SP-related state of distress (Solomonova et al, 2008). One salient feature of felt presence experiences during SP is the fact that it is a distinct sensation, and may occur in the absence of visual, auditory, or tactile imagery. Felt presence experiences during SP have been classified as a paranoid delusion (Cheyne & Girard, 2007), an expression of spatial social imagery (Nielsen, 2007), and as a variant of basic intersubjective experience of the world (Solomonova, Frantova, & Nielsen, 2010).

Felt presence experiences are often interpreted within the cultural framework available to the experiencer (see the following section on the cultural neurophenomenology of SP), but some basic characteristics seem to be common across cultures and ages (Cheyne, 2001): 1) felt presence often manifests from ambiguous stimuli: it is often described as "shadowy", and its physical characteristics are often unclear; 2) the experiencer may report a distinct sensation of being watched, and that the presence has some intentions towards the dreamer; these range from some vague interest to full-blown assault; 3) felt presence is usually accompanied by intense emotions (often fear when the presence is interpreted as threatening), sometimes to the point of a distinct feeling of dread, imminent death, or being in the presence of evil. Positive emotions, however, are also possible, especially when the experience is understood as visitations by deceased relatives or visions of the divine.

## Intruder

Consider the following examples of felt presence experiences of the *Intruder* type: A 26-year-old man from the United States reports: "It felt as if someone was watching me but silently standing behind me". In this example the presence is felt in a distinct and clear way, but not seen or heard, yet the experiencer knows where in space the presence is located. Similarly, a 29-year-old woman from USA regularly experiences the malevolent presence without ever seeing it: "…feeling of evil that is watching or monitoring; never able to actually see this "evil entity". Even in the absence of direct visual, auditory or tactile imagery she feels that she is observed and that the presence is "evil". The ambiguous qualities of the physical attributes of SP visitors can be illustrated by the following two examples. A 39-year-old man from USA writes: "The "presence" is a tall black/darkest grey shadow of a human form without any features.  It stands in the doorway to my bedroom waiting to be "noticed". Likewise, a 30-year-old woman experienced various ways in which the presence was manifesting during her SP attacks: "Once it seemed a shadow was leaving the room. One other time the shadow seemed to have "wild" hair or if it doesn't have hair at all, it looked as some sort of black something".

### Incubus

The *Incubus* experience happens when the *Intruder* physically oppresses the sleeper, sometimes in a rather dramatic way. In words of a 52-year-old man from the United States: "My worst experience was being choked by a man who burst into my bedroom.  The experience was so real and frightening that I was very afraid of my SP for many months after." The *Incubus* takes many forms, including human, supernatural and more rarely, animal: "I often hallucinate creatures like large cats - lions or tigers, … wrapping themselves firmly around me and crushing my body" writes a 20-year-old woman from England.

Some of the most dramatic and potentially traumatic SP *Incubus* experiences are instances that are lived as sexual assault or alien abduction. Consider the following example, reported by a 40-year-old man from the United States: "When it is a "Dark Man" episode, he most likely touches me.  Either by laying across my body, in a sexual way or in the beginning, he would grab me and drag me.  I always felt that if I let go, he would pull me out of my body". Similarly, a 29-year-old woman from Spain describes her distressing SP experience: "…extreme terror, the feeling that air is dense and darker, that shadows boil and take shape… I hear some low tone noises, voices, tactile feeling of grabbing, of naked cold skin, and, very rarely, a presence. Very dark with round eyes, spider-like fingers, that laughs, messes up the bed, and makes me feel terror, with some sexual approaches…". In a study linking reports of space alien abduction to SP episodes McNally and Clancy present this case: "…female abductee… was completely paralyzed, and felt electrical vibrations throughout her body. She was sweating, struggling to breathe, and felt her heart pounding in terror. When she opened her eyes, she saw an insect-like alien being on top of her bed" (McNally & Clancy, 2005: 116).

### Positive felt presence experiences and doubling

While most easily recognizable and most commonly documented cases of felt presence during SP have to do with a threatening and ominous "visitor", some evidence suggests, however, that the presence is not always understood as hostile. Such experiences include perception of friends and family; visitation from deceased relatives or benevolent spirits; and erotic encounters where the sense of presence is comforting. A 20-year-old SP sufferer from the United States writes: "Once or twice I have thought that my friend or roommate was standing over me. I was confused but not afraid." Similarly, encountering deceased family members in visions or in dreams can be experienced as a positive spiritual event, and possibly play a healing role in processes of bereavement (Garfield, 1996; Belicki et al, 2003).

Finally, another rare kind of SP-related felt presence episode involves first an experience of "someone there", and then a doubling of the dreamer's own body, a self-projection into the extracorporeal space. Some individuals report that the felt presence entities are becoming an externalized view of themselves: "Sometimes I feel that the presence is myself, that I can watch myself", reports a 21-year-old man from Jamaica; "I switch to another world and I myself become a presence", writes a 19-year-old man from Russia.

# Body experiences in sleep paralysis

Most (if not all) SP episodes are defined by an altered experience of the body. These include simple experience of muscle paralysis; sensations associated with supernatural assault, including touch, pressure on the chest, or even choking; feelings of unusual vibrations or falling into a vortex; and out-of-body experiences, including flying, falling, or moving around one's house.

One of the most salient features of SP is the REM sleep-related muscle atonia. The inability to move is a striking and unusual experience for most individuals, and the mismatch between sensing the body and the loss of voluntary control over the body's movements may contribute to a range of somatosensory experiences. As discussed above, some of the most intense SP episodes may involve a feeling of being assaulted or touched by a supernatural entity. For instance, a 34-year-old man from USA describes the following experience: "Felt my arms pinned across my chest in a strait jacket hold, felt hands on my chest pinning me against a wall". Perception of not being able to fully breathe, often accompanied by feeling of pressure on the chest, may be prevalent in as much as 57% of SP episodes (Sharpless et al, 2010).

Although most accounts of and research on SP experience have centered on paralysis accompanied by terrifying mentation and by felt presence, not all SP experiences are characterized by imagery and many are simply experiences of transient body paralysis during the transition between sleep and wakefulness, without any other accompanying mental activity (American Academy of Sleep Medicine, 2001). Additionally, SP episodes may be predominantly somatosensory in nature: Cheyne (1999) characterizes these experiences as *Vestibulo-Motor* mentation.

Autoscopy, out-of-body experiences, vibrations, floating, falling and body doubling experiences (Cheyne, 2002) are all possible within the SP framework due to its reliance on dream-supporting REM sleep mechanisms. During a dream, especially a lucid dream (wherein one is aware of the fact that she is dreaming), it is possible to have simultaneous experience of one's dream body and real body at the same time. Thompson (2014) distinguishes between the dreaming self (I the dreamer) and the dream ego (I as dreamt) as two coinciding modes of self-experience, which may sometimes be experienced in parallel. The dreaming self is the sleeping self, it is the "I" of the waking life, now engaging in the practice of sleep and dreaming. The dream ego, on the other hand, is the experiential self, immersed in the dream scenario. The I as dreamt is the temporary "I" that takes on the first-person perspective as a subject (and sometimes an object) of the dream world. Seen from this point of view, SP episodes may represent an intense experience of the dreaming ego, lacking a dream body and temporarily "stuck" within her immobilized sleeping body of the dreaming self/I the dreamer while experiencing dream-like mentation. This feeling of being stuck, coupled with awareness of the overlap between states of vigilance, may then transform itself into a situation of perceptual doubling of body imagery.

Contrary to most SP episodes with a felt presence component, some bodily experiences are described in quite positive terms. For instance, a 20-year-old woman from England describes this characteristic of her typical SP episode: "Generally, the experiences start with a low, pleasant vibration that moves through my body in defined waves, from the feet up. I feel them most strongly in the throat and in my eardrums".

Out-of-body experiences are also relatively common in SP – as much as 39% of SP experiencers have had one at some point (Cheyne, 2002). A 39-year-old woman from the United States writes: "I floated out of my bed into the kitchen. But, as I floated over my bed, I saw like this beast figure crouched over on the front of my bed. I floated over it down to the kitchen. That is where I saw this beautiful kaleidoscope-like leaves. They were so vibrant ... I then floated back to my room into my body". In this example there is a combination of various SP characteristics: dream-like mentation superimposed onto the environment, a nocturnal visitor, and an altered sense of the body.

SP experiences are also sometimes accompanied by false awakenings—dreams where one has a vivid and realistic feeling of waking up in their own bed and engaging in usual activities only to realize that they are still asleep (Buzzi, 2011). While false awakenings are typically characterized as dream experiences, their phenomenology in terms of realism and possible state overlap is to a degree similar to SP. In Cheyne's report (2002), 58% of people who experience SP also experienced false awakenings at least occasionally. Additionally, false awakenings are often associated with feelings of dread, anxiety, and oppression (Green & McCreery, 1994; Nielsen and Zadra, 2011), similarly to SP. The following two examples from our Internet-based sample illustrate such cases: a 24-year old man from the United States reports: "... sometimes I think I have moved... sometimes even gotten up and walked around only to find that I never got up at all." In a similar vein, a 21-year old man from Jamaica describes his experience: "I will wake up into another dream inside my bedroom and think I am awake and realize I am still sleeping minutes later and the same procedure repeats several times".

## Emotions

*...Extreme anxiety and fear, mind is awake, but body is asleep. I feel as though I am trapped and cannot communicate with those around me. (*23-year-old woman, USA*)*

The most prevalent emotion associated with SP experiences is fear. Indeed, the most natural reaction to waking up unable to move is panic, and the sensation of constricted breathing (consistent with REM sleep physiology) may increase the state of distress. Sharpless and colleagues (2010) introduced the term *fearful isolated sleep paralysis* to denote SP experiences characterized by an intense state of distress.

As much as 90% of reported SP episodes are described as fearful (Cheyne, Rueffer, & Newby-Clark, 1999). Similarly, in an Irish University students' sample, fear was found to be the most prevalent emotion, with 82% of respondents stating that they have experienced fear at some point during a SP episode (O'Hanlon, Murphy, & Di Blasi, 2011). Moreover, nightmare frequency was previously reported as a predictor of SP occurrence (Liskova et al, 2016). This data suggests that SP, or at least the *Fearful* form of SP, can be seen as an intensified form of a nightmare: a recent study by Robert and Zadra (2014) reported that about 65% of nightmares and 45% of bad dreams are characterized by fear.

One approach to classify the affective and personal impact of SP experiences is to assess not only frequency or intensity of SP episodes, but also *distress* associated with SP experiences (Solomonova et al, 2008; Cheyne & Pennycook, 2013). To what extent is the individual affected by SP? To what extent do negative emotions carry-over from an SP episode into waking life? Do SP experiences promote a negative relationship with sleep? These questions have been successfully examined in previous research on nightmares (Belicki, 1992; Blagrove, Farmer & Williams, 2004), showing that the individual impact of negative and intense dream experiences depends more on a trait-like reactivity, sometimes referred to as affect distress (Nielsen & Levine, 2007). This trait is thought to represent a general dysfunction of affect regulation network, and it has been shown to be a better measure of how much nightmares influence waking life emotional well-being than frequency or intensity of self-reported nightmare occurrence. Furthermore, affect distress mediates reactivity, negative interpretation and degree of negative reaction to nightmares (Belicki, 1992; Levin & Fireman, 2002). According to Nielsen and Levine (2007; Levin & Nielsen, 2009), dreaming helps regulate emotional memory consolidation and emotional reactivity via fear extinction. Nightmares, therefore, represent a case of problematic/dysfunctional processes of fear extinction. In combination with other factors, affect distress is likely to play a role in formation, experience and interpretation of SP.

Positive emotions associated with SP are much less studied, and it is not possible to accurately estimate their prevalence. One possible reason for this is lack of appropriate screening (SP is often diagnosed as an unpleasant phenomenon) and lack of medical/psychiatric concern: patients are not very likely to describe such experiences to their health practitioner, since they are not bothered by them. In addition, the current diagnostic criteria for a recurrent isolated SP as listed in the latest edition of the *International Classification of Sleep Disorders – 3d edition* (American Academy of Sleep Medicine, 2014), include that the episodes must cause "clinically significant distress including bedtime anxiety or fear of sleep". Such a provision would effectively exclude all possible positive and non-distressing SP phenomena from investigation and/or diagnosis. Nonetheless, in a web-based SP study Cheyne (2002) reports that in addition to anger (30% of respondents) and sadness (23%), bliss (17%) and erotic sensations (17%) are also sometimes present in SP.

## Visual and auditory experiences

Felt presence is the most prevalent, the most emotionally disturbing, and the most salient SP-related experience. Therefore, it is unsurprising that most visual and auditory mention during SP usually has something to do with these unwelcome visitors. The entities, however, while *felt* in a very distinctive and concrete way, are often described visually as rather general and vague shadowy beings. Visual experiences are reported to occur in 54%– 56% and auditory experiences in 55%-60% (Solomonova et al, 2008; Cheyne, 2002) of SP sufferers.

SP may be accompanied by auditory experiences, ranging from abstract and mechanical sounds, such as electric sounds and sounds of buzzing, to vivid auditory imagery, consistent with SP experience of an *Intruder* or an *Incubus*. Sounds of footsteps and of voices are often reported (Cheyne, Rueffer & Newby-Clark, 1999; Cheyne, 2002; Solomonova et al, 2008).

*Cultural grounding of SP*

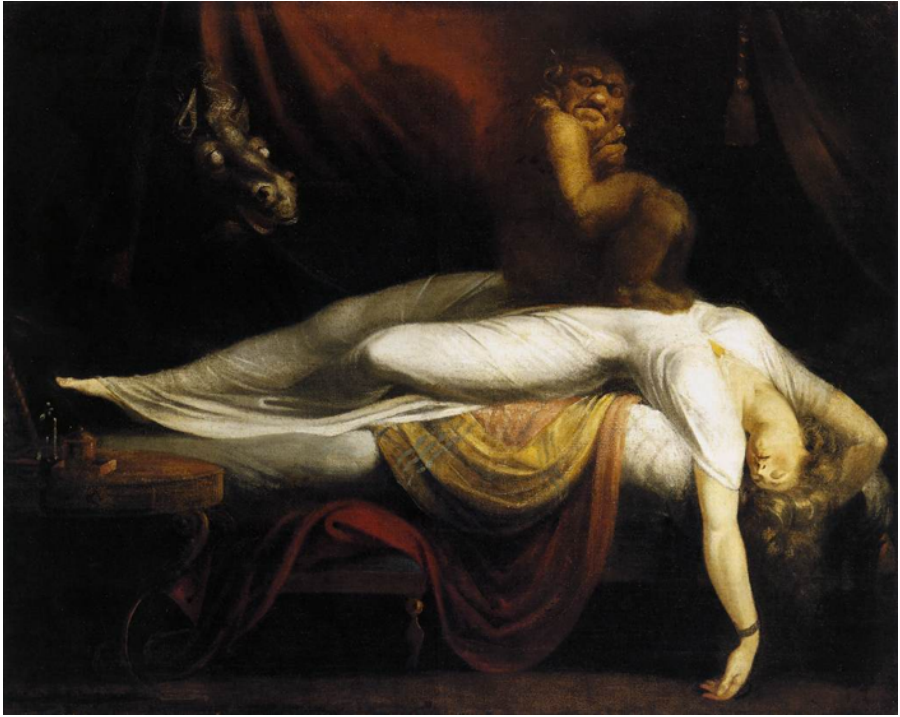

Figure 2. Henry Fuseli. The Nightmare. 1781. Detroit Institute of Art. Public Domain image: https://commons.wikimedia.org/w/index.php?curid=15453518

While sleep paralysis is a lesser-known sleep phenomenon in the West, it is quite prevalent and is well-described in many other cultures. Due to the lack of general awareness of SP in the West, it is rarely discussed in the context of family medicine or psychology. Cross-cultural work on SP revealed that it is rooted in a variety of religious beliefs and cultural schemas, including interpretations of the experience and techniques to engage with the nocturnal visitors. Some of the common qualities of SP across cultures (Adler, 2011) include: 1) sensation of being awake; 2) perception of the environment; 3) paralysis; 4) feeling of fear and dread; 5) felt presence; 6) chest pressure/breathing difficulties; 7) supine position; 6) various unusual body sensations. These apparently culturally-invariant qualities of SP-related occurrences of the experience of a supernatural attack have been at the center of the phenomenological and cross-cultural cognitive research on SP.

Figure 2 is a reproduction of an eighteenth-century work by Henri Fuseli entitled *The Nightmare*. It represents a sleeping woman in a supine position being oppressed by a maleficent creature sitting on her chest and with an ominous presence of the *night-mare*. It is likely that the early use of the English term *nightmare* was to describe intense SP (Orly & Haines, 2014). Culture-specific presentations of SP-related felt presence experiences typically involve a maleficent supernatural being, such as a witch or an evil spirit. Some examples found across cultures include the *kanashibari* demon in Japan (Fukuda et al, 1987; Arikawa, Templer & Brown, 1999); *kokma* in the West Indies (Ness, 1983); "old hag" in Newfoundland (Hufford, 1989); *pandafeche* in Italy (Jalal, Romanelli & Hinton, 2015); *uqumangirniq* among the Inuit of Baffin Island (Law & Kirmayer, 2005); and many others (for a comprehensive list of terms for SP experiences see Adler, 2011). Figure 3 illustrates a possible SP representation (Orly & Haines, 2014): a Japanese demon *Yamachichi* oppresses and inhales the breath of the sleeper.

The first systematic cultural exploration of SP was done by Hufford (1989): he described a phenomenon specific to Newfoundland – the "old hag" witch attack. In his book Hufford discusses the tension in situating SP experiences somewhere between the 'cultural source hypothesis', wherein cultural interpretations and framing influence how an experience unfolds, and the 'experiential source hypothesis', where some invariant lived experiences, such as SP, may influence the development of a spiritual interpretation and formation of cultural beliefs (Hufford 1989, 2005). Similar to this notion, McNamara and Bulkeley (2015) proposed an experiential hypothesis to describe how dreams and other dream-associated experiences, including visions and transcendental experiences, can be seen as a cornerstone and a source of religious belief (McNamara & Bulkeley, 2015). According to this view, a number of cultural, religious and paranormal beliefs are shaped primarily by direct experience and then framed within a particular tradition, which imbues them with existential and metaphysical meaning, a notion that is reminiscent of William James' grounding of mystical experience in the phenomenology of lived experience (James, 1985).

The effect of framing such intense subjective experiences within a cultural tradition can have at least two kinds of potentially opposing effects. On the one hand, many cultures provide not only supernatural explanations of SP, but also remedies and protective rituals against it (some of which are described in the later part of this chapter), thus rooting the SP in a framework which allows for shared narrative and for practical interventions. On the other hand, intense and fearful SP, when interpreted as supernatural assault, has a potential for traumatizing the sleeper, thus creating a vicious circle of anxiety, aversion to sleep, facilitation of future SP episodes (Hinton et al, 2005; Sharpless et al, 2010), and increasing the level of distress via "cultural fear priming" (Ohayon et al, 1999; Jalal, Romanelli, & Hinton, 2015). For instance, the *Incubus* experience, when seen as part of the Christian tradition starting with the late Antique period, according to Gordon (2015), gained additional stigmatizing power, with a connotation of an illicit supernatural sexual experience. Not only were SP victims living through a waking nightmare of an encounter with a demonic assailant, they were also seen as responsible for having summoned it due to their own sinful predisposition/thoughts/impurities.

It is important to note that while SP can include a range of experiences, such as positive experiences, neutral emotions, vestibulo-motor phenomena, out-of-body experiences, and others, most cultural interpretations of SP deal specifically with overlapping aspects of Intruder and Incubus.

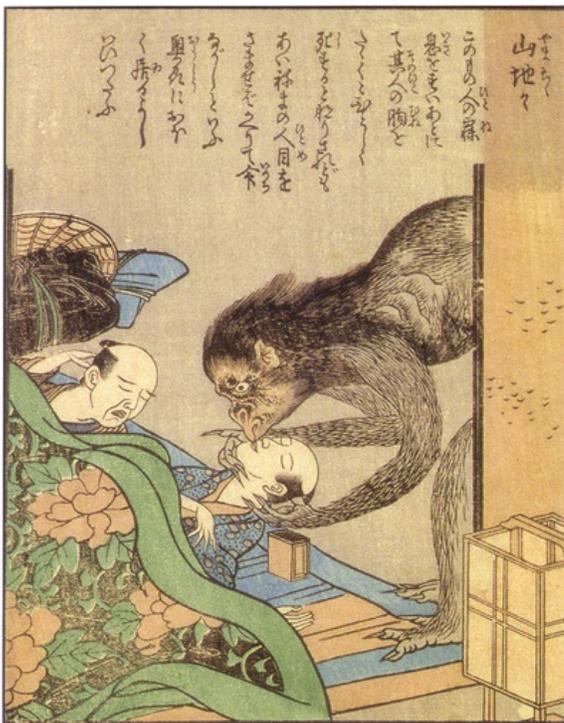

Figure 3. Takehara Shunsen. *Yamachichi*. Public domain image:
https://commons.wikimedia.org/w/index.php?curid=2074508

### Neural basis, associated conditions, and precipitating factors

Human sleep is typically divided into two kinds: REM sleep and NREM sleep. Healthy adults alternate between NREM and REM in cycles lasting about 90 minutes, for a total of 5-6 cycles over a night of sleep.

While it is possible to experience dreaming in all stages of sleep, REM sleep is typically characterized by the most vivid, realistic, bizarre, and emotionally intense sleep mentation (Nielsen, 2000). Other vivid dream experiences, such as nightmares (Nielsen & Levin, 2007) and lucid dreams (LaBerge, Levitan & Dement, 1986), are also typically associated with REM sleep.

Within the context of narcolepsy, sleep paralysis is a part of the diagnostic tetrad, alongside daytime sleepiness, cataplexy, and hypnagogic hallucinations (Thorpy, 2016). There is not sufficient data to assess whether there are significant differences in phenomenology between narcolepsy-associated SP and the isolated form.

SP episodes are characterized by simultaneous presence of waking thought and of REM sleep psychophysiology (Mahowald & Schenck, 1991, 2005; Terzaghi et al., 2012), and the sleeper can often open her eyes and become relatively aware of her environment, while REM sleep-related spontaneous mentation - vivid dreaming - superimposes onto otherwise awake consciousness. This imagery may occur at sleep onset (hypnagogic) or upon awakening (hypnopompic). Other characteristics of REM sleep, such as airway occlusion and rapid shallow respiration (Gould et al., 1988) may contribute to the feeling of being suffocated or the perception of shortness of breath often reported by SP sufferers. Additionally, in one study obstructive sleep apnea was found to be a possible precipitating factor for ISP (Hsieh et al, 2010).

Little research has been done on the sleep characteristics of SP sufferers. Some preliminary data suggests that the SP sleep profile may be similar to that of frequent nightmare sufferers (Nielsen et al, 2010), in that SP participants appear, paradoxically, to exhibit less REM sleep pressure, have more "skipped" REM sleep periods, and show no increase in eye movement density (as opposed to healthy controls) throughout the night (Solomonova, et al, 2012). SP participants also show higher delta power during sleep than non-SP controls (Marquis et al, 2015), which suggests alteration of processes of wake-NREM-REM regulation.

Some of the vestibulo-motor characteristics, such as autoscopy, out-of-body experiences, and feelings of physical transformation, may stem from disturbances in right parietal regions (Jalal & Ramachandran, 2014): the mismatch between intended motor movement and inability to move may contribute to unusual physical sensations.

SP may be experimentally elicited in laboratory settings, but only using an arduous protocol of repeated sleep interruption. For example, SP episodes were experimentally induced by letting participants sleep uninterrupted for the first NREM period, thus eliminating most of the slow-wave sleep pressure (a tendency of slow-wave NREM sleep to take precedence and occupy a large proportion of early night sleep), and then repeatedly awakening participants after 5 minutes of REM sleep have elapsed, thus augmenting REM sleep pressure and facilitating sleep-onset REM periods (SOREMPs). SOREMPs may be seen as a facilitating factor in a REM-wake state dissociation thought to characterize SP experiences. It should be noted, however, that even within such controlled settings and demanding protocols, rates of SP were relatively low: 6 episodes total in 16 participants who already had a tendency toward recurrent isolated SP (Takeuchi et al., 1992), and 8 episodes from 184 sleep interruptions in 13 SP sufferers (Takeuchi et al, 2002). These results suggest that incidence of SP at sleep onset may signify an individual's propensity to enter into REM sleep directly upon falling asleep. This further supports the idea that SP may result from alterations in wake-REM-NREM regulation patterns, resulting in state overlap.

### Associated conditions

Little is known about the epidemiology of SP, but growing evidence points to a combination of genetic and experiential factors. The only study to date to examine genetic factors associated with SP has reported moderate heritability and that this effect was associated with factors known to contribute to disrupted sleep cycles (Denis et al., 2015). Sleep fragmentation and disruption in wake-NREM-REM regulation are an important factor facilitating SP occurrence, but it is uncertain whether all types of SP can be explained by a propensity to sleep fragmentation. Some ethnic groups seem to be more likely to experience SP than others. The Hmong population in Wisconsin, for instance, had a significantly higher incidence of SP than a non-Hmong cohort (Young et al, 2013), with as much as 31% of interviewed Hmong participant reporting at least weekly occurrence of SP episodes. Individuals of African descent also seem to have elevated rates of SP (Bell et al, 1984; Friedman & Paradis, 2002).

Links between affective disorders, especially depression and anxiety, and SP have also been reported. A relationship have been found between SP and depression magnitude and anxiety (Szklo-Coxe, Young, Finn, & Mignot, 2007), social phobia and panic disorder (Paradis & Friedman, 2005; Otto et al, 2006; Sharpless et al, 2010) and social anxiety (Simard & Nielsen, 2005), especially with the sensation of being observed (Solomonova et al, 2008). Changes in REM sleep regulation are often found in mood disorders, especially in depression (Arargun & Cartwright, 2003; Nofzinger et al, 1994)

The relationship between trauma, especially post-traumatic stress disorder (PTSD), and SP has been noted by a number of researchers. McNally and Clancy found that there was a higher proportion of SP reports in participants with a history of childhood sexual abuse (McNally & Clancy, 2005a), and Abrams and colleagues (2008) reported that sexual abuse survivors report more distressing and more frequent SP incidence. In addition, higher rates of SP were found in Hmong population in relation to traumatic Vietnam War experiences (Young et al, 2013), as well as in Khmer (Hinton et al, 2005a) and Cambodian refugees (Hinton et al, 2005b). Similarly, Sharpless and Grom (2013) report that some cases of SP onset in adolescents begin after the loss of a family member. Considering that SP may be conceptualized as a nightmare spectrum experience, this relationship may represent the same dysfunction in the affect regulation network (Levin and Nielsen, 2007; Nielsen & Levin, 2007) as the one that has been proposed to be involved in nightmare production. PTSD-related sleep disturbances have been extensively documented (Spoormaker & Montgomery, 2008; Germain, Buysse & Nofzinger, 2008), including REM sleep dysregulation and increased nightmares (Melman et al, 2002; Germain, 2013), which in itself may contribute to altered REM sleep pressure, in turn facilitating occurrence of SP episodes.

Since SP is often associated with intense, detailed, and troubling visions, a link between SP and psychiatric disorders has been hypothesized. Research, however, shows no consistent relationship between psychiatric conditions and SP, with the exceptions of PTSD, panic disorder, and social anxiety. In one study a number of links between SP and psychiatric conditions were found (Ohayon et al., 1999); these findings,were challenged, however, by an internet-based study (Solomonova et al., 2008), with a larger sample size, in which no strong links between psychopathology and SP were described. However, while isolated SP often presents itself in the absence of psychopathology, higher rates of hypnagogic and hypnopompic experiences (dream experiences occurring during the transition between sleep and wake: at sleep onset or upon awakenings, respectively), some of which may be associated with SP, are often found in psychosis (Plante and Winkelman, 2008).

**Precipitating factors**

In their recent book, Sharpless and Doghramji (2015) list a number of plausible precipitating factors for SP occurrence in susceptible individuals. Sleep fragmentation and insufficient sleep are among the most obvious factors. REM sleep deprivation has been shown to increase REM sleep pressure contributing to REM rebound effect and intensified dreams at sleep onset (Nielsen et al, 2005). Poor sleep quality with frequent awakenings and disruptions may also facilitate REM-wake overlap, creating fruitful conditions for the occurrence of SOREMPs (Takeuchi et al, 1991, 2002; Spanos et al, 1995). Shift work, jet-lag, use of sleep disrupting medication, stress, anxiety – all these factors affect sleep and may facilitate a SP episode. Alcohol consumption was also reported to promote SP (Golzari & Ghabili, 2013; Munezawa et al, 2011), probably due to its effect on altering sleep architecture (Roehrs & Roth, 2001). Sleeping in a supine position also appears to enhance the risk of a SP episode (Sharpless et al, 2010).

**Neurocognitive considerations**
**A return to felt presence**

While undoubtedly felt presences are a hallmark of SP, especially of the intense and frightening episodes, presence experiences are not restricted to this parasomnia and are reported in a variety of conditions, thus possibly representing a more general and basic social imagery process (Nielsen 2007, Solomonova, Frantova & Nielsen, 2011). Arguably, the most salient and compelling felt presence occurs in the context of mystical and spiritual experiences. Otto (1958) introduced the idea of the *numinous* as a cornerstone of religious mystical experiences. Some of the recent work comes from anthropology: the ecstatic presence of God is manifested in the community of Evangelical Christians in the USA (Luhrmann, 2012). Other examples of felt presence have been documented in situations that are physically and emotionally straining or novel. Some examples of these experiences include high altitude climbing (Brugger, Regard, Landis, & Oelz, 1999); feeling of the presence of a baby in postpartum mothers (Nielsen & Paquette, 2007); presence of deceased relatives in the context of bereavement (Simon-Buller, Christopherson & Jones, 1989; Taylor, 2005; Keen, Murray & Payne, 2013); in extreme environments, such as solitary sailing (see also chapter by Suedfeld in this volume), surviving in remote and hostile environments (Suedfeld & Mocellin, 1987), and others. While in most cases felt presence is experienced spontaneously, in some cases it may be a product of sustained mental practices (as in prayer and some forms of meditation). One contemporary non-religious phenomenon is *tulpamancy* (Veissiere, 2016) – a long-term practice of conjuring up imaginary companions, that, over time, may be experienced as almost as real as other people.

Additionally, being able to have a felt sense of others may be seen as a prerequisite for the development of subjectivity. Recent work in phenomenology and enactivism suggests that development of sense of self depends crucially on sensing others, as early as in utero (Gallagher, 2005; Ammaniti & Gallese, 2014), that the sense of one's own body depends on the sense of others (Maclaren, 2008) and that the self-other dynamic is a necessary condition for the sense of self (Zahavi, 2014). Evidence from dream research too suggests that dream processes are relational and intersubjective. The fact that dreams are most often about other people has been conceptualized as simulations of social interaction (Revonsuo, 2016) and as representations of individual attachment styles (McNamara et al, 2001). Additionally, dreams, similarly to waking, can be seen as a dynamic interaction between the "self"-related and "non-self" elements of dream content (everything extraneous to the dreamer). These non-self elements (non-human characters, dream environment, even dream objects can be seen as a "dream other" due to their inherent relational property (Solomonova et al 2015) and to the fact that dream environment in its entirety affectively motivates dreamer to engage with it.
.

**Toward a cultural neurophenomenology of SP**

SP has often been characterized as dissociative (Terzaghi et al., 2012) state, since it effectively combines characteristics of 'waking' consciousness (self-awareness, access to autobiographical memory, ability to open eyes and perceive the environment) with REM-sleep phenomena, specifically muscle atonia/paralysis and mentation/dreams. This notion of SP as dissociative has been at the heart of the previous neurobiological work on the link between dreaming and REM sleep. The relative deactivation of the dorsolateral prefrontal cortex characteristic of REM sleep (Hobson, Stickgold & Pace-Schott, 1988; Maquet, 2000) has been long hypothesized to be at the root of the loss of autobiographic memory and of the inability to appreciate the contents of the dream as "bizarre" or implausible in relation to reality. This has led to the hypothesis that in REM sleep dreaming one is effectively delusional and in a state of a transient psychosis (Hobson, 2004). In SP, similarly, there is often incomplete autobiographical access. This association between SP and REM sleep has also displaced the experience of SP from the psycho-spiritual domain of meaningful encounters with menacing/unreal/supernatural others, into a more reductionist account of uncontrollable and inescapable REM-initiated hallucinations.

In contrast, an account of SP in the context of an oneiric phenomenology and in a 4EA perspective may allow for a more nuanced reading of these experiences. An emerging neurophenomenological framework of sleep challenges strict distinctions between wake, NREM, and REM sleep. Indeed, while SP is one of the examples of simultaneous presence of REM sleep and wake processes, it is not the only phenomenon that attests to the fluidity and interpermeability of states of consciousness. Lucid dreaming is another example of REM-wake co-occurrence (LaBerge, 1986); REM sleep behaviour disorder is characterized by preserved motor output during REM dreaming (Peever, Luppi & Montplaisir, 2014); somnambulism episodes combine NREM and wake physiology and phenomenology (Zadra et al, 2004); and a variety of dream-enacting behaviours, such as laughing, simple movement, crying and looking for a baby in bed, are prevalent in normal populations (Nielsen & Paquette, 2007; Nielsen, Svob & Kuiken, 2010).

A more continuous view of mentation in sleep includes viewing SP as a form of oneiric experience: as a process of intensified mind-wandering (Fox et al, 2013), as a process of creativity (Hartmann & Kunzendorf, 2013), or as enactive imagination (Thompson, 2014), a process of sense-making in a rich, embodied and intersubjective world (Solomonova & Sha, 2016). In his discussion of lucid dreaming, Thompson (2014) proposes that in addition to seeing this state as a dissociative superimposition of two distinct states of consciousness, it may be simultaneously approached as an integrative state, thus allowing for an integration of two different yet related ways of self-experience.

While SP sufferers feel awake and in their own bed, the realism of the experience and the quality of total immersion are completely overpowering to the dreamer, so that she is unable to appreciate the dreamlike quality or the unreality of the SP episode. The high prevalence of tactile and physical sensations probably contributes to this effect. There are, however, numerous accounts of long-time SP experiencers that are characterized by a certain 'feel' for the experience as somewhere between real and unreal. SP-related experiences may have a very compelling and realistic quality, but they are usually lived differently from waking experiences, as a kind of a liminal state.

Consider the following example: while the participant is experiencing intense emotion and is quite absorbed in the unfolding on the SP, he seems to have a kind of a dual awareness regarding the nature of his SP:

> "...*Can't. Move. Not a muscle. Not an eyelash. It's often accompanied by hallucinations. So this bizarre or terrifying event is happening all around me, and I am completely unable to respond or defend myself. Sometimes I know it's not real, somewhere in my mind, but it looks real, and it sounds real, and I'm terrified or revolted (or maybe just bemused), but I cannot wake myself up to stop it.*" (30-year-old man, USA)

Similarly, in another example the experiencer is also hesitant to ascribe any particular state to her experience:

> "... *I might be answering wrong, because I see the beings in my dream-state immediately before waking. But their presence seems so real, I would compare the experience to having them accompanying me in the room*". (48-year-old woman, USA)

Grounding SP in its cultural context allows us to appreciate the variety of factors contributing to qualities of the lived experience, and it may not be possible to dissect the relative contribution of the multitude of neural, phenomenological and cultural narrative factors (Kirmayer, 2009). Importantly, in the current medical context, reducing SP to a dysfunction of REM psychophysiology may also have an important effect on reducing the potential for a deeper exploration of SP as a spiritual experience (Hufford, 2005).

The cultural neurophenomenology of SP is a powerful tool for investigating SP from the 4EA cognition perspective. As neurophysiological, experiential accounts of SP show, the dreamer is in fact embodied – the oneiric scenario is dependent on the dreamer's state of consciousness (REM intrusion) and on the dreamer's physiological state (atonia, shallow rapid breathing). She is embedded in a physical (interprets ambiguous stimuli around her) and in a cultural world (these ambiguous stimuli take on a familiar shape/are infused with a deeper cultural and interpersonal signification). The sleeper is also extended into the world – the whole environment, both dreamt and real, is part of her ongoing experience; and her experience is enactive – there is a relational quality: she is not a passive observer of the oneiric drama unfolding before her eyes, but rather she is deeply engaged (Solomonova & Sha, 2016).

In order to elucidate neurophenomenological qualities of SP in greater detail, future work may use microdynamic phenomenology/elicitation interviews, aimed at uncovering the fine-grained temporal and structural qualities of lived experience (Nielsen, this volume; Petitmengin, 2006; Petitmengin & Lachaux, 2013), in addition to neurophysiological data and deep awareness of the cultural, religious and spiritual context of the experiencer.

### *Sleep paralysis practices: prevention, disruption, treatment and exploration*

While SP remains a relatively unknown phenomenon in much of Europe and America, a number of practical culture-specific practices have been developed to protect the sleeper from the negative influence of presumed supernatural forces. While some of these methods have deep roots in their respective metaphysical contexts, and therefore need to be grounded in existing religious and mystical practices, a number of practical and conceptually neutral recommendations have emerged, and seem beneficial for most SP sufferers, regardless of background.

No established treatment for SP currently exists; its clinical management is instead often focused on treating comorbid problems. According to a review by Sharpless & Doghramji (2015), psychoanalysis, cognitive-behavioural therapy (CBT), hypnosis, and education in sleep hygiene have been investigated in relation to SP, but no empirical consensus on efficacy of such interventions is currently available. Based on the available evidence on SP and cognitive-behavioural approaches to treatment of sleep disorders, especially insomnia, the authors propose a manual for CBT-ISP. This is a promising first step toward finding a systematic method of dealing with SP. Sparse evidence for pharmacological interventions for SP also exists: in one study it was suggested that REM sleep-suppressing antidepressants may provide temporary relief (Plante and Winkelman, 2008), and treatment of narcolepsy may reduce SP frequency (Mamelak et al., 2004).

Antidepressants and anxiolytics were also used in severe cases (Hsieh et al., 2010). Terrillon and Marques-Bonham (2001) proposed that management of SP might benefit from administration of melatonin, which would help normalize the circadian rhythm. The cost of side effects associated with these treatments, however, may outweigh the benefit, and Shapless and Doghramji (2015) argue for a cautious approach, tailored to each individual situation.

While methods for dealing with sleep paralysis have not been systematically explored by empirical psychology or cognitive science, the contemporary context of Internet-facilitated support groups and information sharing practices are changing the solitary and culture-bound nature of SP attacks. Furthermore, a number of methods have been anecdotally reported and documented online and in print, that see SP experiences as an opportunity rather than a nuisance, and promote exploration of one's own consciousness via SP-supported lucid dreaming or even contemplative approaches to SP (Hurd, 2010). One popular support group-mailing list is known as "Awareness during sleep paralysis" (ASP), and a reddit group on SP counts over 4000 users, sharing information on the phenomenology of their experiences and methods of overcoming them.

Cultural and clinical practices associated with SP can be roughly separated into three kinds: 1) preventative practices, focused on avoiding SP-enabling circumstances; 2) disruptive practices, designed to stop SP in the middle of the experience; and 3) observational/explorative practices, aiming at observing SP and possibly transforming it into a positive event, such as a lucid dream or an out-of-body experience.

Raising awareness of SP-associated phenomena itself may be one of the most important factors in reducing fear and distress before, during and after SP occurrence (Otto et al., 2006; Sharpless et al, 2010). Indeed, knowing that the experience is transient (will not last), benign (does not contain any real danger), and common (is shared with many individuals across the world) are powerful tools for psychological distancing and for facilitating an eventual observational, as opposed to fully immersive and fatalistic, attitude toward SP. Knowing about SP phenomenology and neurophysiology may have access to cultural grounding with available symbolic gestures helps prevent, disrupt and transform a negative experience into a tool for self-exploration. Figure 4 summarizes the intricate links between precipitating factors and effects of SP experiences in light of disruptive, and observational/transformational practices.

## Methods for preventing sleep paralysis

While undoubtedly helpful, simply knowing the basis of SP may not be enough to alleviate terror and distress associated with the experiences, and disruption techniques are clearly warranted. A 25-year-old man from the United States reports: "This happens sometimes every night, sometimes only once every few weeks. Even though I 'know' what is happening, and that I am in no danger, it is always terrifying". The first study to systematically assess prevention strategies for SP by Sharpless and Grom (2014) has suggested that while no foolproof method for preventing SP is yet known, some strategies, such as avoiding sleeping on one's back (supine position), maintaining optimal sleep hygiene (avoiding stimulants, noise, irregular sleep patterns and anything that contributes to sleep fragmentation), and pre-sleep relaxation practices may help in preventing SP.

A number of culture-specific preventative ritualistic measures to prevent SP exist. These include placing a variety of defensive objects in the room or in the bed before going to sleep, such as a variety of knives (Hufford, 1982, Law & Kirmayer, 2005); sprinkling salt (a common anti-witch remedy) (Roberts, 1998); putting a broom bottom-up (Paradis & Friedman, 2005) or a pile of sand at the bedroom door (Jalal, Romanelli & Hinton, 2015); and many others. Putting a Bible in the room (Hufford, 1982) and saying a protective prayer before bedtime are also thought of as effective deterrents. Other ritualistic actions, designed to deter, divert and chase away unwelcome supernatural visitors were also documented in a variety of contexts (Sharpless & Doghramji, 2015).

## Techniques for disrupting sleep paralysis

While preventative measures, whether culturally embedded or aimed at increased awareness and promotion of sleep hygiene, may be effective in reducing the frequency of SP episodes, many methods for dealing with an ongoing SP experience also exist. Considering that most SP experiences are characterized by fear and other unpleasant sensations, it is not surprising that in one study the majority of participants reported having attempted to disrupt the ongoing SP experience. Moving the extremities and self-monitoring (raising awareness, promoting calm) may be helpful during the SP episode (Sharpless & Grom, 2014). Not all attempts or all strategies are equally successful, but it seems that attempting micro movements, instead of trying to get up or to scream, are most effective. Culture-bound rituals include saying a prayer (Hufford, 1982), making a sign of a cross with one's tongue (Davies, 2010), and asking someone to physically shake the oppressed sleeper (Law and Kirmayer, 2005).

### Observational/transformational practices

One may argue that ISP and lucid dreaming are polar opposites. However, they share the same underlying psychophysiology and seem to involve similar mechanisms: both are dependent upon REM sleep mechanisms; both are characterized by simultaneous presence of the dream state and by the feeling of being awake, including activation of higher order metacognitive functions indicative of some degree of waking thought processes (LaBerge, Levitan & Dement, 1986; Voss, Holzmann, Tuin, & Hobson, 2009; Dresler et al, 2012; Filevich et al, 2015); and in both cases muscle atonia is present. The crucial difference between the two states is the quality and the focus of awareness and metacognition: in lucid dreaming one is aware of the illusory nature of the dream scenario, whereas in SP the dreamer is often absorbed by the vision, not always fully realizing that it is dreamlike, and, in case of fearful SP, is too absorbed in the panicky state of perceived imminent danger.

The link between SP and lucid dreaming has not been systematically investigated in empirical research, but two studies report a positive correlation between frequency of lucid dreaming and SP (Denis & Poerio, 2016; Solomonova, Nielsen & Stenstrom, 2009), suggesting that the REM-wake intertwined state, characterizing SP, may be a trait-like phenomenon predisposing individuals to SP on the one hand, and facilitating lucidity in REM sleep dreams on the other.

Transforming SP into a positive experience, such as an OBE or a lucid dream, or utilizing SP experiences as a means of contemplative insight into one's own mind, may become a practice in itself, since not only techniques for disrupting and preventing SP exist in the contemporary digital culture, but also techniques for inducing SP, with the hope that the experience will function as a portal to a desirable altered state of consciousness (Hurd, 2010). The following two reports illustrate the transformative potential of SP:

> *"I have woken up from dreaming and found I can't move or open my eyes. I get the feeling of lemonade bubbling in my body, especially my head. It is very frightening. But since I have been having OBE[3]s I now relax and go with the flow of sleep paralysis and sometimes I actually achieve an OBE"* (40-year-old man, Australia)

> *"At first I was very frightened until I found the ASP email group and found that I was not the only one being "visited" by this being during sleep paralysis. ... When it first started happening it was more of an assault and I had to fight terribly to escape. But after years, I learned to ignore and now I've been trying to communicate with the presence". (*40-year-old man, USA)

---

[3] OBE = out-of-body experience

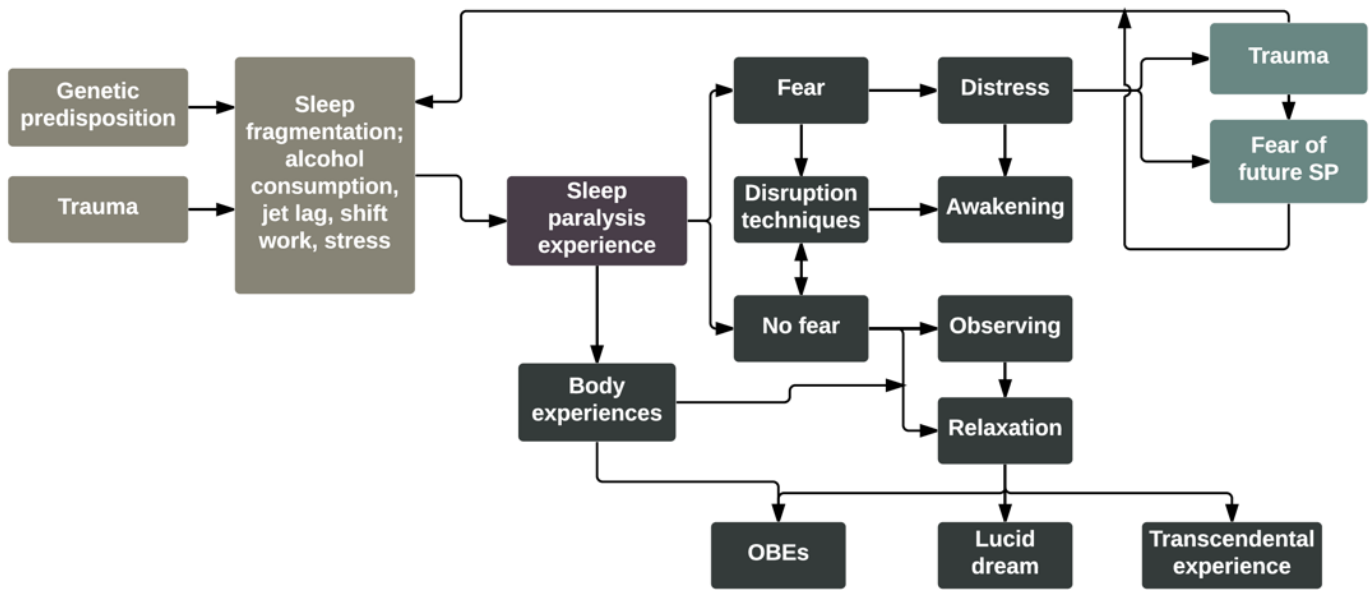

Figure 4. Predisposing, precipitating factors and experience and outcome of sleep paralysis episodes.

### *Further considerations and future directions*

In terms of possible avenues for treatment, since SP can be conceptualized as a form of nightmare occurring in a mixed state of consciousness, nightmare treatment techniques could be useful in approaching SP. Currently, the most used and recommended technique for treating chronic nightmares is the Imagery Rehearsal Therapy (Krakow et al, 1995; Krakow & Zadra, 2006), which consists of "rehearsing" and transforming dysphoric oneiric imagery in a safe context. This method has been effective in treating PTSD-related nightmares (Krakow et al, 2001; Germain et al, 2004; Cook et al, 2010; Casement & Swanson, 2012), which seems particularly appropriate for intense and trauma-related SP experiences. Similarly, treatment of nightmares by lucid dreaming is a promising avenue (Zadra & Pihl, 1997; Spoormaker & Van Den Bout, 2006; LaBerge, 2009). Considering that neurophysiologically both states are characterized by an overlap between REM sleep and wakefulness, and that a number of folk approaches treating SP as a portal to lucid dreams already exist, mastering lucid dreaming could be an effective approach to transformation of an ongoing SP episode. Such a strategy may also be highly effective in de-stigmatizing and desensitising the experiencer, and especially in increasing her mastery and agency over her spontaneous oneiric experiences.

Contemplative practices, such as meditation or pranayama (yogic breathing) may also be useful in dealing with recurring SP episodes. There is currently no empirical evidence for contemplative techniques and SP management, with the exception of a case study by Jalal (2016), but anecdotal evidence from practitioners as well as growing empirical literature linking contemplative practices with stress management, emotion regulation, and increased self-awareness, provide grounds for future research.

Recent years have seen an important increase in empirical studies on the effects of meditation and meditation-based mindfulness interventions. There are documented benefits of contemplative practice in clinical populations including positive effects in mood disorders such as anxiety and depression (Hoffman et al, 2010; Goyal et al, 2014), social anxiety (Goldin & Gross, 2010), and PTSD (Kearney et al, 2013). At least four kinds of meditation are currently investigated in relation to mental health: focused attention, open monitoring (Lutz et al, 2008), self-transcendence (Travis & Shear, 2010) and loving kindness meditation (Hoffman, Grossman & Hinton, 2011). Different kinds of meditation practices may recruit different neural networks (Fox et al, 2016), and particular psychological and neuroplastic changes, associated with meditation practice, likely depend on the kind and duration of meditation experience (Lutz et al, 2015). These different kinds of contemplative practice may be helpful in targeting different kinds of recurrent SP experiences, promoting de-automatization (Kang, Gruber & Gray, 2013): deconstructing patterns of behaviour/reactivity.

Meditation may be effective in SP management as a way of cultivating a non-judgemental or 'non-sticky' observational attitude to arising imagery, sensations and emotions, and in letting the experience unfold. In addition, one important feature of most mindfulness-related practices is the focus on the experience of the body (Kerr et al, 2013), and some evidence suggests that meditation practice may improve awareness of one's own body states (Solomonova et al, 2016) and increase introspective accuracy for somatic experience (Fox et al, 2012). Breathing practices, such as pranayama, may be particularly effective in transforming SP as it is happening due to the fact that many SP episodes are characterised by a feeling of disordered/insufficient breathing. A recent study (Seppälä et al, 2014) reported that breathing exercises were effective in decreasing PTSD symptoms in war veterans. This implies that practicing techniques that improve awareness of body sensations may lower the reactivity to SP episodes, thus lowering the distressing quality of the experience, and increasing the potential for disrupting or transforming SP.

**Acknowledgements:**

The author was supported by Social Sciences and Humanities Research Council (SSHRC) of Canada and by a J.-A. DeSèves Sacre-Coeur Hospital Foundation doctoral scholarship. Thanks are due to Tore Nielsen, Philippe Stenstrom and Michelle Carr for numerous conversations on sleep paralysis and its interpretation, and to the members of the Dream and Nightmare Laboratory at the Center for Advanced Research in Sleep Medicine. Additional thanks are due to Don Donderi and Elena Frantova, as well as to all participants who have consented to share their experiences with us. Special thanks to Benjamen Samaha (artist) and to the anonymous SP sufferer for generously offering the narrative and the drawing of a sleep paralysis episode.